# An IoT based Real-time Low Cost Smart Energy Meter Monitoring System using Android Application

Md Redwanul Islam[1], Supriya Sarker[2], Md Shahraduan Mazumder[3], Mehnaj Rahman Ranim[4]

1,3,4(Department of Computer Science & Engineering, World University of Bangladesh, and Dhaka, Bangladesh
Email: radwan143@gmail.com[1], mjmajumder55@gmail.com[3], mehnajislam2016@gmail.com[4])

2(Department of Computer Science & Engineering, Chittagong University of Engineering & Technology, and Chittagong, Bangladesh
Email: sarkersupriya7@gmail.com)

**Abstract:**

Nowadays IoT based applications are becoming more popular because it provides efficient solutions for many real time problems. In this paper, an IoT based electric meter monitoring system using android application has been proposed that aims to reduce manual efforts for measuring the electricity units and make users concern about the excessive usage of electricity. Aurdino Uno and an optical sensor are used to fetch the pulse of the electric meter. In order to reduce human error and cost in energy consumption, a low cost wireless sensor network is implemented for digital energy meter and a mobile application that automatically capable of interpret the units meter.

*Keywords* — **IoT, Electric Meter, Optical sensor, mobile application, cloud server, android, DPDC,**

## I. INTRODUCTION

The advancement of the Internet of Things has been emerging day by day. The Internet of Things (IoT) revolves connection between M2M that embedded with electronics, software, sensors, actuators that assist users in monitoring and controlling devices remotely and efficiently [1]. In the IoT based system object and living being are provided with unique identifiers with the ability to transfer data [2]. The area of IoT has amplified from the convergence of wireless technologies, microelectromechanical systems and the Internet [2]. Nowadays IoT technology is being applied in many areas like electricity, gas, water etc. to make our life automated. Nowadays due to the excessive use of the internet, these areas become computerized and online payment system makes possible. But accessing meter reading is a manual process and has the possibility of error which causes high revenue cost [3]. Automatic Meter Reading (ARM) technology facilitates the assessment of energy consumption and analysis of data for billing and payment. ARM technology requires to bring the device online and connecting device with the internet which is in other term Internet of Things [1]. ARM technology using wireless communication is cheaper than wired medium. Hence, WiFi is more suitable for the proposed system as it is very common in every residence.

According to statista, the number of smartphone users in the world was 2.87 billion in 2017 [4]. The percentage of android phone user in Bangladesh was 79.21% in December 2017 [5]. This statistics indicates usage of Android phones in the country. Hence, we choose the Android platform to implement our system. The total amount of installed electricity generation capacity (including captive power) in Bangladesh is 15,351 megawatts (MW) estimated in January 2017. In 2015, 92% of the urban population and 67% of the rural population have access to the facility of electricity [6]. People have to check their electricity meter reading manually. In Bangladesh, there are some people who don't understand meter reading. Hence, they do not know the used amount of electricity.

The aim of the work is to design and develop a real-time low-cost energy meter monitoring systems integrated with cloud services along with an android application.

The rest of the paper is organized as follows. Section II reviews the related papers aligned with the proposed system. Section III explains the model of the proposed system. Section IV illustrates the results and analysis. Section V concludes with limitation and future directions.

## II. LITERATURE REVIEW

In recent years enormous research and papers have proposed the design and development of energy meter monitoring system. In [7] the author proposed a Wi-Fi based single-phase smart meter based on IoT. The author used a digital meter, ESP8266 Wi-Fi module and a web application for the user interface. The ESP8266 Wi-Fi module has attached into the meter. The ESP8266 Wi-Fi module has been implemented by TCP/IP protocol as the means of communication between the meter and web application. The proposed system is secured and open source but costly. In [8] an Automatic Meter Reading (ARM) based Power Meter with Wi-Fi Communication Module scheme has been proposed and developed software based on Linux. In [9] the author presented a survey report on the utilization of smart electricity meters and some key aspects of the metering process. As well as opportunities arising due to the advent of big data and the increasing popularity of the cloud environments challenges are highlighted by the author. In [10] the author proposed a system where Arduino Uno has been developed with an Ethernet shield that can monitor all the necessary activities in the flow of electricity, the use of current and electricity costs and mentioned hope of reducing the problems associated with payment, calculate the cost of the unit of electricity. In [3] the author proposed a real-time monitoring system for residential energy meter. The presented system provided inclusive and continuous access to energy consumption to the consumer by exploiting the advancement of IoT technology.

## III. PROPOSED MODTHODOLOGY

This section discusses the architecture and working principles of the proposed energy meter monitoring system. It also shows the device prototype and interconnection of components while working in real-time.

### A. Architecture

The architecture of the proposed system is divided into three major parts. They are Energy meter region of the customer, Cloud server region and Application interface for the customer.

*1. Energy meter region of customer*

This region consists of two electric meters, two devices and a Wi-Fi network. The devices have been connected with the Wi-Fi network to transfer data in the cloud server. The devices are being mounted on the energy meter to fetch the pulses. The received pulses are being transferred to the corresponding accounts of customers created in the cloud server. The customers need to be registered using the Android application [11] as well as login their account and can request to access all of their information regarding energy consumption.

*2. Cloud Server region*

A database is created into the cloud server for each device. Each device fetches pulses from the energy meter which is converted to kilowatt using the equation (1) and transferred into the database (FIREBASE) [12]. Besides, Chip ID, Hostname, IP address, Mac address, pulse count, the total amount of consumed power are stored in the database.

*3. Application Interface for customer*

The users can view all stored data in the cloud through an android application. Users can request information on their registered accounts by log into corresponding accounts. To login, the user has to input user id and password as a means of security. After login, the user can view the number of consumed units, received pulses, total cost, details of the Wi-Fi network. A help manual guide the users to use the application. In the case of pre-paid, users set an initial balance of recharged money. The application automatically calculates the number of consumed units and notifies the users when 80% of the balance of recharged money has been paid.

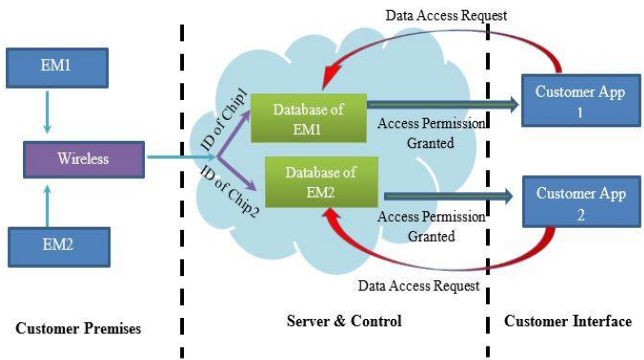

**Fig. 1: Proposed Model**

### B. Flow of the Process

Initially, the device is being mounted on an energy meter. The device has to install with the IP address of the nearby wireless network to connect with the internet. If the information of wireless network is saved beforehand, the device will be connected to the wireless network automatically. Through the wireless network, the device is being connected to the cloud server and checks for any previous information in the database. If it does, then start counting from the next, otherwise, it starts from zero.

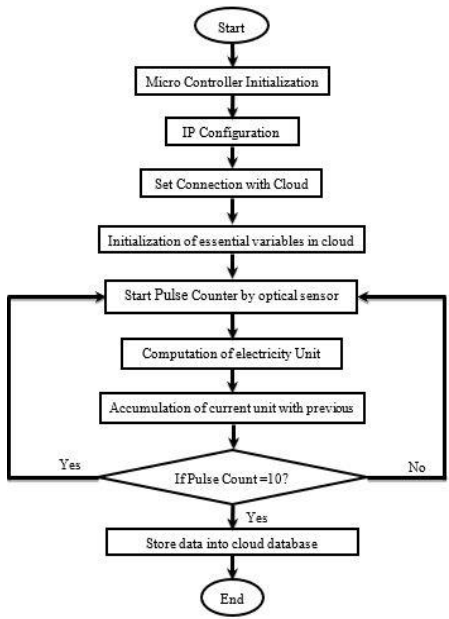

**Fig. 2: Flow of Proposed Model**

The device contains an Arduino Uno [13] microcontroller where an optical light sensor, an RGB LED light and a booster is being connected. The booster helps to boost the battery volts. The battery is used as an alternative power source. The optical light sensor calculates pulse from the electric meter and converts it to the readable data. Those data will be stored on the specific ID of the cloud after the specified time. The users can request the cloud through the android application and retrieved results accordingly.

### C. Device Prototype

The customer region of the developed system has implemented in real time in residence sector. The device has been mounted on the energy meter.

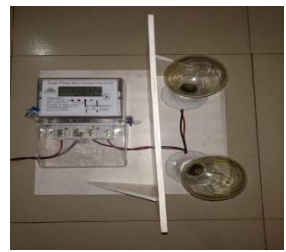 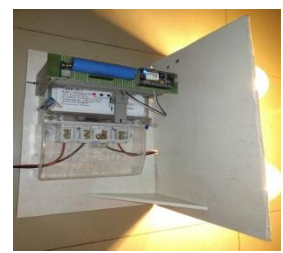

**Fig 3(a): Device is offline**   **Fig 3(b): Device is online**

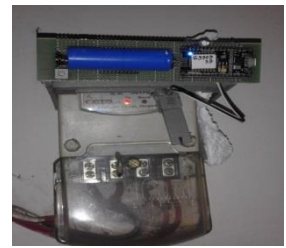 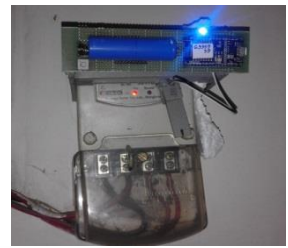

**Fig 3(c): Device before receiving pulses**   **Fig 3(d): Device while receiving pulses**

In fig 3(a) the device is mounted on the energy meter and two bulbs have been set up. Power is not supplied here. In fig 3(b) the system has been connected to a continuous power source and two bulbs are consuming energy. While the optical sensor receives a pulse from energy meter the LED light is put on (in fig 3c), otherwise LED light remains off (in fig 3d).

## IV. RESULTS & ANALYSIS

In this section, we are going to discuss the process of calculating the amount of consumed power, calculation of pre-paid and post-paid charge. The pulse output may be a flashing LED or a switching relay (usually solid state) or both. The Node MCU use to collect pulse from the meter through the optical sensor. In case of an energy meter, a pulse output can be defined by an amount of energy passing through the meter [14]. Dhaka Power Distribution Company Ltd (DPDC) sets some specific unit ranges. The amount of bill is calculated according to the cost of each unit for both prepaid and post-paid meter [15]. Our developed android application followed the pre-defined rules set by DPDC while calculating the consumed amount and payable bill.

### D. Pulse Calculation

For single-phase, domestic electricity meters (e.g. the Elster A100c) each pulse usually equals 1Wh (1000 pulses per kWh) 1000 pulses = 1KWh

$$1 \ pulse = \frac{1}{1000} kwh \quad (1)$$

### E. Unit Calculation of Energy Meter

1) *Unit Calculation for Pre-paid meter:*

Let Total cost = T
$$T = T_R + D + V \quad (2)$$
where $T_R = Total\ cost\ range$
$$= R_1 + R_2 + R_3 + R_4 + R_5 + R_6 \quad (3)$$

1st range $R_1 = U_1 * pu1$
2nd range $R_2 = U_2 * pu2$
3rd range $R_3 = U_3 * pu3$
4th range $R_4 = U_4 * pu4$
5th range $R_5 = U_5 * pu5$
2nd range $R_6 = U_6 * pu6$
where U1,U2,..U6 are units of range from 1 to 6
pu1, pu2, …pu6 are per units of consumption from range 1 to 6

$$D = demand\ charge = 150\ BDT \quad (4)$$
$$V = VAT = 5\%\ of\ (T_R + D) \quad (5)$$

2) *Unit Calculation for Post-paid meter:*

Let, Total Purchased Energy= E
$$E = A - (V + M_R + D) \quad (6)$$
where,
A= Paid amount
$M_R$= Meter rent = 40BTD
D = Demand Charge = 50 BDT
V= Vat = 5% of (A-$M_R$)
R= Rebate = 1% of A-V+D
Total Unit Consumption,
$$U = U_1 + U_2 + U_3 + U_4 + U_5 + U_6 \quad (7)$$

### F. Database View

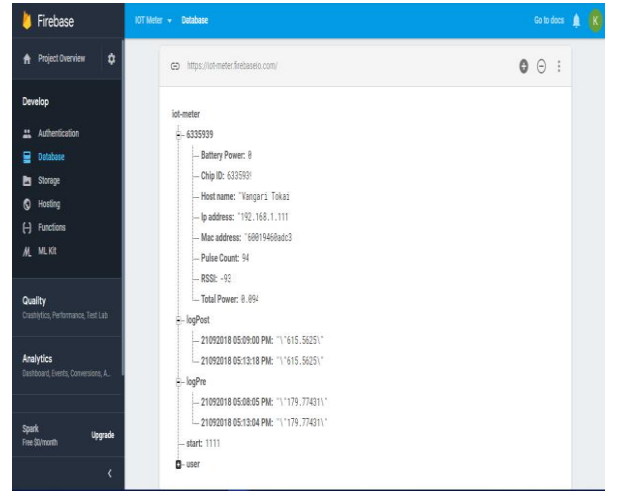

**Fig 4: View of Database in the Cloud (Firebase)**

In Fig 4, it is shown that the parameter of the energy meter, wireless network, and consumed units for both prepaid and post-paid meter has been stored.

### G. Application Interface

This section demonstrates some pages of the developed application interface. The users need to be registered before using the application for the first time providing the corresponding device ID. After that, they can log into their account using the password (fig 5.a) and request their meter data to the cloud database.

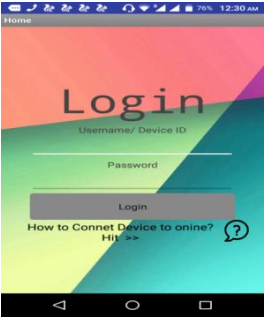
Fig 5(a): Login Interface

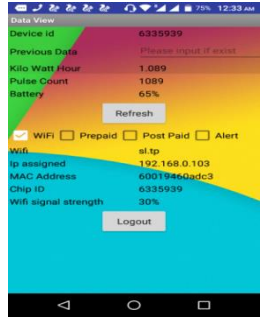
Fig 5(b): Wifi Information

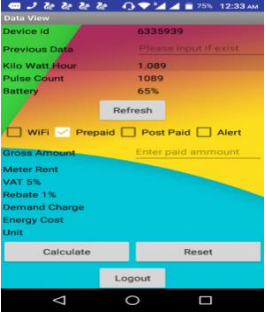
Fig 5(c): Unit calculation for Prepaid

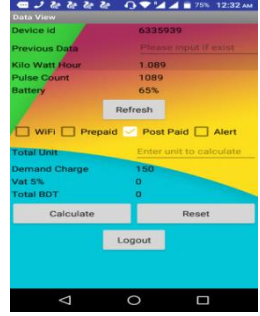
Fig 5(d): Unit calculation for Post paid

User can view their WiFi information (in fig 5b), unit calculation according to DPDC rules both for pre-paid (in fig 5c) and post-paid (in fig 5d) depending on their requests.

### H. Experimental data

We test the proposed system in real-time to measure the accuracy on it. We recorded data of energy consumption in 7 consecutive days. The time difference between each collected data is 24 hours. The developed module has been connected to an energy meter to observe the energy consumption pattern of that house.

| Days | Pulse Count | Kilo Watt | VAT (5%) | Total (BTD) |
|------|-------------|-----------|----------|-------------|
| Day1 | 14208 | 14.2 | 10.34/= | 217.14/= |
| Day2 | 15600 | 15.6 | 10.62/= | 223.02/= |
| Day3 | 15336 | 15.3 | 10.56/= | 221.76/= |
| Day4 | 16440 | 16.4 | 10.78/= | 226.38/= |
| Day5 | 13944 | 13.9 | 10.28/= | 215.88/= |
| Day6 | 14664 | 14.7 | 10.44/= | 219.24/= |
| Day7 | 16200 | 16.2 | 10.74/= | 225.54/= |

**Table 1: Electricity Bill Calculation**

Table I shows the list of recorded pulses in seven consecutive days of a certain week. Moreover, the amount of energy in KW, vat and total cost of each day is calculated in the table.

## V. DISCUSSION

The optical sensor receives a pulse in every 6 seconds from energy meter and converts it to data; 10 pulses of the electric meter are being transferred and stored in a database periodically. For that reason, it is instructed to transfer 10 pulses from the device at a time to the cloud database considering the time difference.

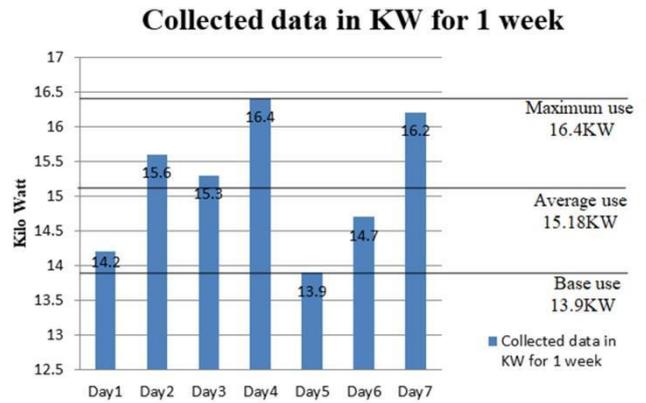

**Fig. 6: Energy consumption in 7 consecutive days**

Fig.6 depicts the amount of consumed energy in seven consecutive days of a week. The vertical axis implies seven consecutive days of a week and the horizontal axis implies the amount of consumed energy (in kilowatt) in a respective day of that week. The system has recorded 14.2 KW, 15.6 KW, 15.3 KW, 16.4 KW, 13.9 KW, 14.7 KW, 16.2 KW energy consumption in Day 1, Day 2, Day 3, Day 4, Day 5, Day 6, and Day 7. We classify the amount of energy consumption in three stages: (1) Base use, (2) Average use, (3) Maximum use. Base use is the minimal amount of usage which means it is not possible to reduce the use of electricity below the amount in a certain week. Average use is the average of base and maximum use. From the data collected by the proposed system, the average use of the entire week is noted as 15.18 KW. The base and the maximum usage of electricity are 13.9KW and 16.4KW, respectively. Analysing the recorded data, it is clear that the user should follow the energy consumption way of Day 5 to maintain minimal consumption or average uses to reduce the power consumption. Users can notice the day of their maximum use and abide from that usage pattern. This graphical chart helps to make concern of excessive use of energy in their residence. Besides, close observation of collected data can be used to find abnormal consumption patterns and defects in our used electrical devices at home.

## VI. CONCLUSIONS

The proposed system is able to reduce the sufferings of the customer and make users concern about the excessive consumption of electricity as well as faulty devices at home. Through this system, customers can easily view total pulse, total units and total costs of electricity. The system is easily readable and reliable. The data stored in the cloud has great importance in future energy meter data mining. In a large sense, energy distribution company such as DPDC able to observe the pattern of consumption of an area. Consequently, this observation can help in load distribution in a certain area.